\documentclass[twocolumn,pra,superscriptaddress,amsmath]{revtex4-2}
\usepackage{graphicx}
\usepackage{hyperref} 

\begin{document}
\title{Tuning the universality class of phase transitions by feedback: \\ open quantum systems beyond dissipation}
\date{July 6, 2021}

\author{D. A. Ivanov}
\affiliation{Department of Physics, St. Petersburg State University, St. Petersburg, Russia}
\author{T. Yu. Ivanova}
\affiliation{Department of Physics, St. Petersburg State University, St. Petersburg, Russia}
\author{S. F. Caballero-Benitez}
\affiliation{Instituto de F\'isica, LSCSC-LANMAC, Universidad Nacional Aut\'onoma de M\'exico, Ciudad de M\'exico 04510, M\'exico }
\author{I. B. Mekhov}
\email{Igor.B.Mekhov@gmail.com}
\affiliation{Department of Physics, St. Petersburg State University, St. Petersburg, Russia}

\begin{abstract}
We shift the paradigm of feedback control from the control of quantum states to the control of phase transitions in quantum systems. We show that feedback allows tuning the universality class of phase transitions via modifying its critical exponent. We expand our previous treatment [D. A. Ivanov, T. Yu. Ivanova,  S. F. Caballero-Benitez, and I. B. Mekhov, \href{http://dx.doi.org/10.1103/PhysRevLett.124.010603}{Phys. Rev. Lett. {\bf 124}, 010603 (2020)}] of Dicke model and go beyond the approximation of adiabatically eliminated light field. Both linearized and nonlinear models of spin ensembles are considered. The tunability of quantum fluctuations near the critical point by the feedbacks of nontrivial shapes is explained by considering the fluctuation spectra and the system behavior at single quantum trajectories.     
\end{abstract}

\maketitle

\section{Introduction}

In our letter \cite{prl2020}, we have studied feedback-induced phase transitions (FPT) \cite{prl2020,SciRep2020,Mazzucchi2016Opt} in open quantum systems, which are not dissipative ones, but are coupled to a measurement device. The first experiment implementing the feedback control of Dicke phase transition using a Bose--Einstein condensate (BEC) trapped in a cavity has been reported in Ref. \cite{EsslingerFeedback}. The general concept  of quantum phase transitions (QPT) \cite{SachdevBook} is very important not only in physics of various systems (e.g. atomic and condensed matter ones), but in other fields as well, e.g., quantum information and technologies \cite{Osterloh}, machine learning \cite{Nieuwenburg}, and complex networks \cite{Halu2013}. In contrast to familiar thermal transitions, QPT is driven by quantum fluctuations existing even at zero temperature in closed systems. Since the extensive studies of lasers \cite{Haken-book} the notion of phase transitions has been adopted to non-equilibrium open systems. The dissipation in open systems provides fluctuations via the system-bath coupling, and the dissipative phase transition (DPT) takes place between nontrivial non-equilibrium steady states \cite{Kessler2012,Daley}. 

In Ref. \cite{prl2020} we extended the consideration of phase transitions in open systems by including the quantum measurements and feedback control. The quantum description of a system being continuously measured is more detailed than that of a dissipative system: the latter is its limiting case, where the measurement results are completely ignored \cite{Wiseman}. We have shown that adding the measurement-based feedback can induce phase transitions. Moreover, the feedback allows for controlling quantum properties of the transition by tuning its critical exponents, and thus their universality class \cite{prl2020,SciRep2020}. In the perspective of such feedback-induced phase transitions the role of quantum fluctuations becomes especially emphasized. The quantum noise here drives the phase transition similar to QPT for closed systems, but it becomes now ``better observable'' via the stochastic nature of the measurement outcome. Moreover, the effect of quantum noise is now twofold: the direct disturbance of the system via the measurement backaction and the feedback loop enhanced fluctuations of the measured signal. 

Such tunable feedback-induced phase transitions can be obtained in many-body problems, e.g., the feedback-induced supersolid-like, antiferromagnetic, and other states in optical lattices \cite{Mazzucchi2016Opt,Spielman2021}, time crystals \cite{prl2020,buonaiuto2021dynamical}, etc. The influence of quantum measurements on phase transitions in many-body systems was considered without feedback, e.g., in Refs. \cite{Mazzucchi2016PRA, Kozlowski2016PRAnH, UedaCrit2016, AshidaNComm2017, AshidaAdvPhys2021,FisherPRB2019,PhysRevX.11.011030, muller2021measurementinduced, zhang2021universal, block2021measurementinduced, minato2021fate, doggen2021generalized, PhysRevLett.126.170602, PhysRevB.103.224210, PhysRevX.9.031009, PhysRevB.98.205136, PhysRevX.10.041020, PhysRevLett.125.030505,PhysRevB.101.104302, PRXQuantum.2.010352, PhysRevB.102.054302, Lavasani}.  

Feedback is a very general idea of modifying system parameters depending on the measurement outcomes: the examples spread from the contemporary rock and classical music (e.g. the Sampo device \cite{Sampo}) to reinforcement learning \cite{Petruccione}. The concept of feedback control has been successfully extended to quantum domain \cite{Wiseman,HammererRMP, HaukePRA2013,HuardFB2016, Hacohen2016, HarocheBook,Mabuchi2004, Sherson2015,Sherson2016, Hush2013, SzigetiPRA2009, SzigetiPRA2010, Bouchoule2017,Hammerer2016PRA, Thomsen2002,St-K2012,Vuletic2007,Wallentowitz2004} resulting in quantum metrology aiming to stabilize nontrivial quantum states and squeeze their noise \cite{Ivanov_2014,jetp2015,Ivanov_2016}. The measurement backaction typically defines the limit of control, thus, playing an important but negative role. In Ref. \cite{prl2020} we have shown that the  feedback can do more than just the quantum state control, but it allows for the phase transition control as well. In this case the measurement fluctuations and feedback enhanced fluctuations drive the transition playing an essentially positive role.

In its essence the quantum system coupled to a classical feedback loop is a hybrid system. Such systems are actively studied in the field of quantum technologies, where various systems have been already coupled \cite{Kurizki2015}: atomic, photonic, superconducting, mechanical, etc. The goal is to use advantages of various components, while avoiding using their disadvantages. In this sense, we address a hybrid quantum-classical system, where the quantum system can be a simple one providing the quantum coherence, while all other properties necessary for tunable phase transition are provided by the classical feedback loop: nonlinear interaction, non-Markovianity, and fluctuations. 

We have shown \cite{prl2020} that FPT leads to effects similar to particle-bath problems (e.g. spin-boson, Kondo, Caldeira-Leggett, quantum Browninan motion, dissipative Dicke models) describing very different physical systems from quantum magnets to cold atoms \cite{Iftikhar,breuer2002,Leggett1987,LeHur2008, DomokosPRL2015, DomokosPRA2016,Scarlatella2016,Plenio2011}. In particular it becomes possible to address important questions about quantum-classical mapping between Floquet time crystals \cite{Sacha2017,Eckardt2017} and long-range interacting spin chains. Our model is directly applicable to many-body systems such as those of many-body cavity QED (cf. for reviews \cite{Mekhov2012,ritsch2013,Review2021,defenu2021longrange}). These systems have been recently marked by experimental demonstrations of superradiant Dicke \cite{EsslingerNat2010, EsslingerFeedback}, lattice supersolid \cite{EsslingerNature2016, Hemmerich2015}, and other phase transitions and phenomena \cite{LevPRL2018, Zimmermann2018,Bux2013,Brantut2020}, as well as theory proposals \cite{Caballero2015,Caballero2015a, Rogers2014,Morigi2010,Niedenzu2013, Gopalakrishnan2009,Kollath2016,piazza2015,Diehl2013,DomokosPRL2015,DomokosPRA2016}. Nevertheless, effects we discuss here require to go beyond the cavity-induced autonomous feedback \cite{St-K2017}, where the feedback time response is limited to an exponentially decaying function.

Here we provide a more detailed theoretical description of the feedback control of phase transitions focusing on the Dicke model. We go beyond the approximation of the adiabatically eliminated light field \cite{prl2020} and treat light as a fully dynamical variable. Interestingly, while the adiabatic model corresponds well to the experiments \cite{EsslingerNat2010, EsslingerNature2016,EsslingerFeedback}, the non-adiabatic model corresponds to the experiments \cite{Hemmerich2015,Kessler2014, HemmerichScience2012}. We consider both linearized and nonlinear models of spin ensembles. The tunability of quantum fluctuations near the critical point by the feedbacks of nontrivial shapes is explained by considering the fluctuation spectra and the system behavior at single quantum trajectories.

\section{Dicke model with feedback}

\begin{figure}
\centering
\includegraphics[width=0.95\linewidth]{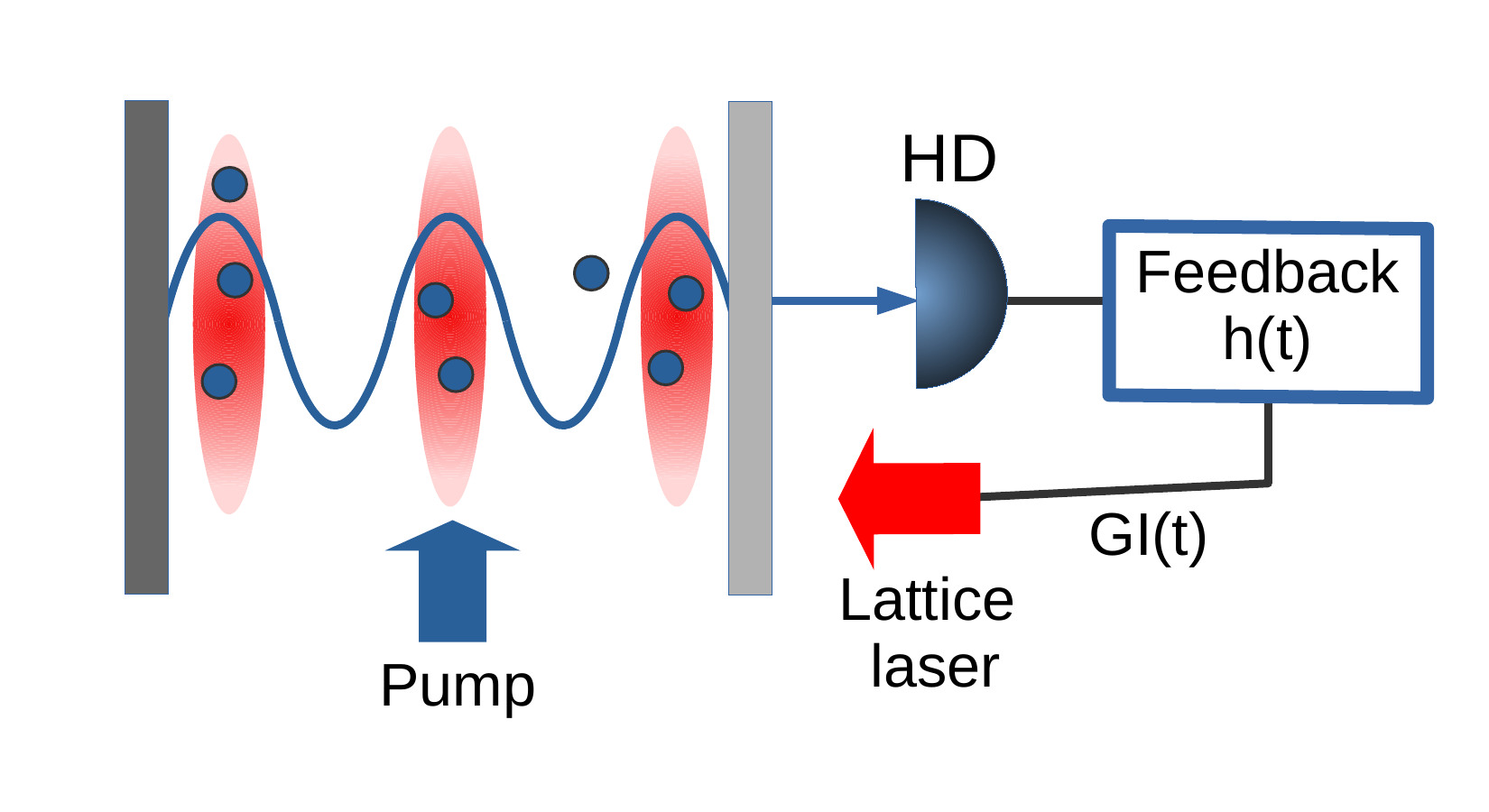}
\caption{\label{figsetupbec} A possible realization of the feedback-controlled Dicke model with a BEC. The atoms are trapped in a 1D periodic potential (red) with the potential depth controlled by changing the intensity of the lattice laser. Additionally, the atoms are coupled to two light modes: the transverse pump and a cavity mode (blue). The cavity light is measured by the homodyne detector (HD). The signal of the detector is processed with the kernel $h(t)$ and used to feedback control the lattice potential.}
\end{figure}

In order to illustrate new features of a phase transition offered by the feedback control we consider the Dicke model with an additional feedback loop \cite{prl2020}. Let us consider a collection of $N$ spins (two-level atoms, qubits) equally coupled to a single bosonic (optical cavity) mode. If we denote the cavity frequency as $\delta$ and the spin frequency as $\omega_{R}$ then the system Hamiltonian becomes
\begin{equation}
\label{eq:SB_H_no_feedback}
H =  \delta a^\dagger a + \omega_R S_z + \frac{2 g}{\sqrt{N}} S_x \left(a^\dagger + a\right).
\end{equation}
Here $a^\dagger$ and $a$ are the creation and annihilation operators of the bosonic mode and $S_i = \sum_j \sigma_{i}^{(j)}$ are the components ($i = \{x,y,z\}$) of the collective spin operator with $\sigma_{i}^{(j)}$ being the Pauli matrices of $j$-th spin. The coupling strength between the spins and the bosonic mode is characterized by the coupling constant $g$. The units with $\hbar = 1$ are used throughout the paper.
      
It is well-known~\cite{Keeling_AQT2018} that the open Dicke model possesses the superradiant phase transition that separates the normal phase with the number of quanta in the boson mode independent from the number of spins $N$ and the superradiant phase where the number of quanta scales with $N$. In the absence of feedback the critical value of the coupling constant reads~\cite{Keeling_AQT2018} 
\begin{equation}
\label{eq:critical}
g_\mathrm{crit} = \sqrt{\frac{\omega_R \left(\kappa^2 + \delta^2 \right)}{4 \delta}} ,
\end{equation}
where $\kappa$ is the cavity decay rate. The direct observation of the superradiant phase transition for two-level atoms in a cavity is impractical due to low values of the atom-field coupling. Nevertheless the transition has been observed with internal atomic states coupled via Raman transition \cite{Dimer_PRA2007,Zhiqiang_Optica2017} and with momentum states of a Bose-Einstein condensate in an optical cavity \cite{EsslingerNat2010,Hemmerich2015}. In principle, the feedback control can be implemented with both these models. Moreover, other systems with the Hamiltonian similar to Eq.  (\ref{eq:SB_H_no_feedback}) can be proposed and equipped with a feedback loop \cite{prl2020}.

The critical exponents for the Dicke model has been found to be different for open and closed systems. In particular, the critical exponent $\alpha$ that describes the divergence of the number of photons in a cavity ($\sim |g - g_{crit}|^{-\alpha}$) equals $1/2$ for a closed system and 1 for an open one. This difference demonstrates the effect of the environment-induced fluctuations on the properties of the phase transition. It was shown \cite{DomokosPRL2015,DomokosPRA2016,LundgrenPRA2020} that the appropriate environment engineering can gradually change the critical exponent. However, experimental implementation of required spectral functions is usually a difficult task: in a BEC the atomic dissipation is limited by the Beliaev damping \cite{Domokos2018}. In contrast, the feedback that we describe here can easily be implemented. We will demonstrate that the feedback control of phase transitions can easily modify the critical behavior of the system.   

For concreteness we consider an implementation of the feedback control in the Dicke model based on the momentum states of a BEC \cite{DomokosPRL2010,EsslingerNat2010,Hemmerich2015}. We assume that the BEC is trapped in a quasi-1D periodic potential $V(t)\cos^2(k_0 x)$, where $V(t)$ is the amplitude of this potential that is controlled by the feedback loop, see Fig. \ref{figsetupbec}, $k_0$ is the wavenumber of the laser forming such a lattice.

In addition to this periodic potential, the BEC is coupled to an optical cavity. The modes of this cavity are labeled as $a_l$, the frequencies of the modes are $\omega_l$. Note that we do not require the high finesse of this cavity. Contrary to the experiments \cite{EsslingerNat2010,Hemmerich2015} in the feedback scheme the cavity can be considered as a quantum meter for the atoms or spins. Thus it would be even beneficial to have a cavity that produces only little effect on the atoms. Moreover, having enough sensitivity of light detectors the feedback scheme can possibly be realized without a cavity \cite{SciRep2020}. The Hamiltonian of the atoms and the cavity modes reads
\begin{equation}
\label{eq:H-BEC}
H = \sum_l \omega_l a_l^\dagger a_l + \int_{0}^{L} \Psi^\dagger(x) H_\mathrm{a1} \Psi(x) dx
 + \sum_l \zeta_l \left(a_l^\dagger + a_l\right), 
\end{equation} 
where $\zeta_l$ are the external classical pumps of the light modes. Here the single-atom Hamiltonian $H_\mathrm{a1}$ after adiabatic elimination of the excited atomic level is written as
\begin{gather}
H_\mathrm{a1} = \frac{p^2}{2 m_a}  + V(t) \cos^2 (k_0 x) \nonumber\\
+\frac{1}{\Delta_a} \sum_{l, m} g_l u_l^*(x) a_l^\dagger g_m u_m(x) a_m ,
\label{eq:H-single-atom}
\end{gather}
where the first term is the atomic kinetic energy with $m_a$ being the mass of the atom, and $\Delta_a$ is the detuning between the mode frequency and the atom transition frequency. The coupling between the atoms and the light modes are described by the constants $g_{l,m}$, the mode functions are $u_{l,m}(x)$.

The simplest configuration allowing for the detection of the distribution of atoms along the lattice axis involves only two light modes, see Fig. \ref{figsetupbec}. One mode is the transverse pump with the uniform distribution along the lattice axis. The corresponding mode function can be chose as $u_\mathrm{pump}(x) = 1$. The other mode has the spatial dependence $u_1(x) \sim \cos k_1 x$. Assuming the frequency of the two modes equals to the double frequency of the lattice laser the scattering of the light from the pump mode $u_\mathrm{pump}$ into the mode $u_1$ will be maximal if the atoms will be tightly localize in the lattice sites. Indeed the localized atoms will serve as a diffraction grating with the period $d = \lambda_0/2$ providing the diffraction maximum condition if $\lambda_0 /2  = \lambda_1$. Thus the uniformity of the BEC distribution determines the amount of light scattered from $a_\mathrm{pump}$ into the mode $a_1$. This light can be used for the feedback control.

Let us decompose the field operator according to the spatial dependence of the Hamiltonian
\begin{equation}
\label{eq:psi-decompose}
\Psi (x) = \frac{1}{\sqrt{L}} c_0 + \sqrt{\frac{2}{L}}c_1 \cos k_1 x,
\end{equation} 
where $c_{0,1}$ are the annihilation operators of the atomic waves with momenta $0$ and $k_1$, respectively. After the substitution of Eqs. (\ref{eq:H-single-atom}) and (\ref{eq:psi-decompose}) in Eq. (\ref{eq:H-BEC}) the Hamiltonian transforms to
\begin{equation}
\label{eq:H-BEC-Dicke}
H =  \delta a^\dagger a + \omega_R S_z + \frac{2}{\sqrt{N}} S_x \left[ g\left(a^\dagger + a\right) + G I(t)\right],
\end{equation}
which is the Hamiltonian of the Dicke model (\ref{eq:SB_H_no_feedback}) with the feedback control part proportional to $I(t)$. In Eq. (\ref{eq:H-BEC-Dicke}) we introduced the notations: $\delta = \omega_1-\omega_0 + Ng_1^2/(2\Delta_a)$ is the detuning of the pump mode and the measured mode including the dispersive shift, $\omega_{R} = k_1^2/2 m_a$ is the recoil frequency, the effective coupling rate is $g = \Omega_p g_0 \sqrt(N/2)/\Delta_a$ with the pump Rabi frequency $\Omega_p = g_\mathrm{pump} a_\mathrm{pump}$. The feedback is performed via $GI(t) = \sqrt{N/8} V(t)$. The spin operators are $S_x = (c_1^\dagger c_0 + c_0^\dagger c_1)/2$, $S_y = (c_1^\dagger c_0 - c_0^\dagger c_1)/2i$ and $S_z = (c_1^\dagger c_1 - c_0^\dagger c_0)/2$. The total number of atoms $N = c_1^\dagger c_1 + c_0^\dagger c_0$ is assumed to be fixed and therefore can be considered as a c-number. The obtained Hamiltonian (\ref{eq:H-BEC-Dicke}) describes the Dicke model with the possibility to control the lattice potential.   

If the number of atoms $N$ in a BEC is large, and the excitation number is small, the Hamiltonian in Eq. (\ref{eq:H-BEC-Dicke}) can be linearized and approximated as \cite{prl2020}
\begin{equation}
\label{eq:H-BEC-Dicke-lin}
H =  \delta a^\dagger a + \frac{\omega_{R}}{2} c_1^\dagger c_1 + \left(c_1^\dagger + c_1 \right) \left[ g\left(a^\dagger + a\right) + G I(t)\right].
\end{equation}

Let us now specify the feedback control. The purpose of the control is to steer the atoms towards the phase transition and modify the critical behavior. Thus the natural algorithm is
\begin{equation}
\label{eq:fb-algorithm}
I(t) \sim \int h(t-\tau) \left(a + a^\dagger \right) d\tau,
\end{equation}
which acts on the atoms in a similar way as the cavity field. The new feature due to the feedback is the time-response kernel $h(t)$ that can be customized to affect the critical behavior of the system. Thus the signal for the feedback loop should be obtained from the measurement of the cavity mode quadrature. Equation~(\ref{eq:fb-algorithm}) is the simplest possibility that illustrates the purpose of the feedback control. Below we will define a more general feedback algorithm that will be demonstrated to have certain advantages over Eq.~(\ref{eq:fb-algorithm}).

The practical realization of the feedback loop in the system with a BEC should not possess technical difficulties. The characteristic frequency of the system dynamics is $\omega_{R}$ which for the rubidium BEC is $\omega_{R} = 2\pi\cdot$ 4kHz. The bandwidth of modern electronics approaches GHz range. Such a bandwidth allows for almost any interesting feedback regimes including the quasi-instantaneous feedback.

One can think of other implementations of the feedback loop as well. A possible option would be the control of the coupling constant $g$, for example, via the pump mode $a_\mathrm{pump}$ as has been implemented in the experiment \cite{EsslingerFeedback}. This would result in highly nonlinear dynamics. 

Another option is, instead of creating the external potential $V$ with $k_0=k_1/2$, one can inject the feedback signal directly through the cavity mirror as the pump $\zeta_1(t)$. In this case, an additional laser with doubled wavelength is not necessary. While the initial Hamiltonian Eq. (\ref{eq:H-BEC-Dicke}) will be somewhat different, after the adiabatic elimination of light mode (as in Ref. \cite{prl2020}), the equation for the spin components will be the same as in the case we consider here, if one chooses the pump through the mirror as $\zeta_1(t)\sim I(t)$.

Note that the notion of the collective effective spins in ultracold gases is not limited to the momentum modes. In optical lattices two effective modes can be represented by atoms in two spatial modes (positions). The simplest case of two spatial modes is atoms at odd and even sites of an optical lattice. The phase transition in such a lattice system leads to the so-called lattice supersolid state, which has been obtained experimentally \cite{EsslingerNature2016,Hemmerich2015}. The physical origin of both types of phase transitions (with and without a lattice) can be traced to the spatial self-organization of cold atoms into a checkerboard pattern in a transversely pumped cavity \cite{ritsch2013}. An important property of such lattice systems is that they can be generalized to many spatial modes \cite{Elliott2015} (modeling large spins) and include real spin degrees of freedom of ultracold fermions. The effective spin operators will then be represented by sums of on-site atom number operators $\hat{n}_i$: $S_x\sim \sum_i A^n_i \hat{n}_i$ \cite{Mekhov2012,MekhovLP2010,MekhovLP2011,Kozlowski2016PRAnH}, or the so-called bond operators $b_i$ (i.e. the annihilation operators of an atom at a lattice site $i$, $\hat{n}_i=b^\dag_i b_i$) representing the matter-wave interference between neighboring sites: $S_x\sim \sum_i A^b_i b^\dag_i b_{i+1}$ \cite{Kozlowski2015PRA, Kozlowski2017, Caballero2015, CaballeroNJP2016} (here, $A^n_i$ and $A^b_i$ are some coefficients tunable by the optical geometry). For example, the effective spin can correspond to the atom number difference between odd and even sites [for $A^n_i=(-1)^i$] \cite{Mazzucchi2016PRA,Mazzucchi2016NJP}, represent the magnetization \cite{Mazzucchi2016PRA} or staggered magnetization \cite{Mazzucchi2016SciRep} of fermions, etc.

We have already demonstrated \cite{Mazzucchi2016Opt} the stabilization of many-body states by feedback in such lattice systems (e.g. the feedback-generated supersolid-like and antiferromagnetic states), which is an example of the feedback-induced phase transition (FPT) as well, as it appears above some critical feedback strength. Expanding our idea of tuning the universality class towards truly many-body interacting atomic and molecular \cite{MekhovLP2013} systems can be a promising direction of research.

\section{Heisenber-Langevin approach and the linearized model}

We start the theoretical analysis of the system by deriving the Heisenberg-Langevin equations with feedback~\cite{Wiseman}. The dynamics of the spins inside the leaky cavity can be obtained from the Hamiltonian~(\ref{eq:SB_H_no_feedback}) adding cavity decay with the rate $\kappa$ and vacuum noise $f_{a}$ entering the system via the partially transparent mirror. The vacuum noise correlation function is $\langle f_a(t+\tau) f_a^\dagger(t)\rangle = 2\kappa\delta(\tau)$. The resulting Heisenberg-Langevin equations read
\begin{eqnarray} 
\label{eqn:HLE-NL} 
\dot{a} &=& -i \delta a - i \frac{2 g}{\sqrt{N}} S_{x} -\kappa a + f_a, \nonumber \\
\dot{S}_{x} &=& - \omega_{R} S_{y}, \nonumber \\
\dot{S}_{y} &=& \omega_{R} S_{x} - \frac{2}{\sqrt{N}} [g(a^{\dagger} + a) + GI(t)] S_{z}, \nonumber \\
\dot{S}_{z} &=& \frac{2}{\sqrt{N}} [g(a^{\dagger} + a) + GI(t)] S_{y}.
\end{eqnarray}
The added feedback term in Eq.~(\ref{eqn:HLE-NL}) drives the collective spin with the strength proportional to the feedback control signal $I(t)$. In order to write the feedback contribution in the similar form as the direct coupling term the feedback gain $G$ is explicitly separated. The feedback control $I(t)$ is an operator in the Heisenberg picture. The expression for  $I(t)$ that is even more general than that given by Eq. (\ref{eq:fb-algorithm}) is
\begin{equation}
\label{eq:quadrature}
I(t) = \sqrt{2\kappa} \int_{-\infty}^{\infty} h(t-z) \mathcal{F} \left[\xi_\theta(z)\right] dz.
\end{equation}
Here the kernel $h(t)$ encodes the timing of the feedback algorithm and the function $\mathcal{F}$ describes the instantaneous transformation of the measured homodyne signal $\xi_\theta(z)$ by the feedback loop. The simplest possibility that will be considered below is the linear feedback with $\mathcal{F}[\xi_\theta(z)] = \xi_\theta(z)$. 

Since the homodyne measurement is performed on the light outside the cavity the vacuum noise reflected from the output mirror contributes to the signal. This effect is taken into account using the input-output formalism \cite{GardinerZoller}
\begin{equation}
\label{eq:input-output}
\xi_\theta(t) = \sqrt{2\kappa}x_\theta(t) - f_\theta/\sqrt{2\kappa},
\end{equation}
where $x_\theta=(a e^{-i\theta}+a^\dag e^{i\theta})/2$ is the light quadrature and the Langevin force $f_\theta = (f_a e^{-i\theta} + f_a^\dagger e^{i\theta})/2$ is the quadrature of the vacuum noise $f_a$.  

The analysis of nonlinear operator equations~(\ref{eqn:HLE-NL}) is rather difficult. It is however possible to estimate the critical condition, where the non-zero spin projection $\langle S_x \rangle$ emerges. Taking the averages in Eqs.~(\ref{eqn:HLE-NL}), neglecting the quantum correlations and quantum noise terms, assuming that the system below transition is weakly excited (approximating $\langle S_z \rangle$ as $-N/2$), and assuming the stationary solutions with zero derivatives, one obtains for the critical feedback gain
\begin{equation}
\label{eq:G-crit-nl}
G_{\rm crit} H(0) = \frac{\omega_R(\kappa^2 + \delta^2) - 4\delta g^2}{4g\kappa C_\theta},
\end{equation}
where $H(0) = \int_0^\infty h(t) dt$ is the zero-frequency Fourier coefficient of the feedback kernel $h(t)$ and $C_\theta = \delta\cos\theta + \kappa\sin\theta$. Note that using the feedback one can reach the phase transition even for the coupling constant $g < g_\mathrm{crit}$ that is below the critical value given in Eq.(\ref{eq:critical}).

\begin{figure}
\centering
\includegraphics[width=0.95\linewidth]{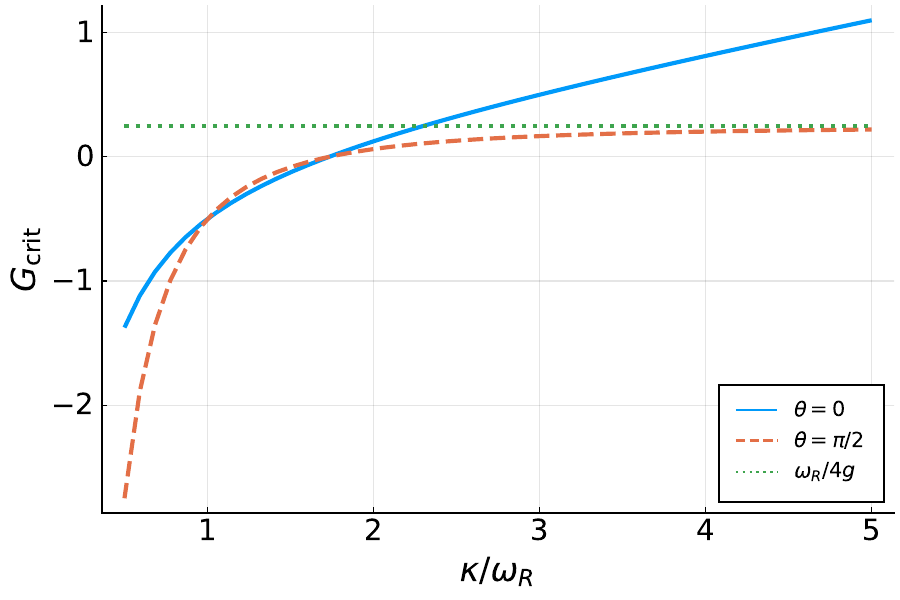}
\caption{\label{fig:Gcrit-kappa}Dependence of the critical value of the feedback gain on the measurement strength $\kappa$ for the quadratures $x_{0}$ ($x$-quadrature) and $x_{\pi/2}$ ($y$-quadrature). If $x$-quadrature of the cavity field is measured, $G_\mathrm{crit}$ grows for large $\kappa$. If $y$-quadrature is measured, then $G_\mathrm{crit}$ approaches the constant value $\omega_{R}/(4g)$. For small $\kappa$ the critical gain becomes negative since the phase transition here happens even without feedback. Other relevant parameters are $g = \omega_R$, $\delta = \omega_R$,  $h(0)=s$.}
\end{figure}

The dependence of critical value of the feedback gain on the cavity decay rate $\kappa$ is shown in Fig~\ref{fig:Gcrit-kappa}. The trend for large $\kappa$ depends on the measured field quadrature. If $x$-quadrature is measured (dashed curve), that is $\theta = 0$, then the critical value $G_\mathrm{crit}$ linearly increases with $\kappa$. Using this regime for an experimental implementation is not very convenient. The optical cavity in feedback experiments is an auxiliary component helping to increase measured signal. However choosing to measure $x$-quadrature one should limit the value of $\kappa$ (or use strong feedback) taking a sufficiently high-fines cavity. On the other hand, measuring the $y$-quadrature there will be no restrictions on the value of $\kappa$: the larger $\kappa$ the closer $G_\mathrm{crit}$ to its limiting value $\omega_{R}/(4g)$. Thus it is advantageous to  use the $y$-quadrature measurement. This type of measurements will be assumed considering all the numerical examples in this paper. Interestingly, within the approximation of the adiabatically eliminated light presented in Ref. \cite{prl2020}, changing the quadrature, which is measured, affects only some coefficients of the final model. In contrast, in the full model without adiabatic elimination, the measurement of different quadratures can lead to rather different equations.  

Note that for small values of $\kappa$ the critical gain $G_\mathrm{crit}$ becomes negative for both measurement scenarios. This indicates that the phase transition in these cases takes place even without any feedback. Another perspective would be to use negative feedback  $G < 0$ to prevent the transition to superradiant phase and keep the system in the normal phase even for $g > g_{crit}$.  This may be important for enhancing the quantum properties of light in the field of cavity QED of quantum gases \cite{MekhovNaturePh2007,Mekhov2012,ritsch2013,Review2021}. Moreover, this will be useful for obtaining fully quantum optical lattices \cite{Caballero2015,Caballero2015a,CaballeroNJP2016,SantiagoFermi2017, CaballeroPRA2016, lozanomendez2020spin} with a plethora of novel phenomena unobtainable in standard optical lattices created by prescribed classical laser beams. As we have shown for such systems \cite{Caballero2015,CaballeroPRA2016}, the quantum properties of light are the most important, when they are not masked by strong light scattering associated with the self-organized phase.    

Let us now perform the linearization and focus on the quantum behavior of the system below the phase transition. Using in the Hamiltonian the Holstein-Primakoff representation $S_z = b^\dagger b -N/2$, $S_{-} = \sqrt{N-b^\dagger b} b$, $S_{+} = b^\dagger \sqrt{N-b^\dagger b}$, $S_x = (S_{+} + S_{-})/2$, and then assuming the number of excitations being much smaller than $N$, the linearized Langevin-Heisenberg equations can be found. In terms of spin quadratures these equations take the following form
\begin{eqnarray}
\label{eq:HL-linear}
\dot{X} &=& \omega_R Y , \nonumber \\
\dot{Y} &=& -\omega_{R} X - 2gx - G I, \nonumber \\
\dot{x} &=& \delta y - \kappa x + f_x , \nonumber \\
\dot{y} &=& - \delta x -\kappa y -2gX + f_y.
\end{eqnarray}
Here $X = \left(b^\dagger + b\right)/2$ and $Y = \left(b - b^\dagger\right)/2i$ are the atom-field quadratute operators, while $x = \left(a^\dagger + a\right)/2$ and $y = \left(a - a^\dagger\right)/2i$  are their cavity-field analogues. The quadrature Langevin forces are defined as $f_x = \left(f_a^\dagger + f_a\right)/2$ and $f_y = \left(f_a - f_a^\dagger\right)/2i$. The feedback signal operator in quadrature notation reads
\begin{equation}
\label{ea:I-operator}
I(t) = 2\kappa \int_{-\infty}^{t} h(t-z) \left[x_\theta(z) - f_\theta(z)/(2\kappa)\right] dz.
\end{equation} 
The linear operator equations~(\ref{eq:HL-linear}) can be solved using Fourier transformation. The algebraic system of equations for the Fourier amplitudes takes the form
\begin{gather}
\label{eq:FT-matrix}
\begin{bmatrix}
\kappa +i\omega & -\delta & 0 & 0 \\
\delta & \kappa + i\omega & 2g & 0 \\
0  & 0  & i\omega & -\omega_R  \\
2g + 2\kappa GH(\omega)\cos\theta & 2\kappa GH(\omega)\sin\theta & \omega_R & i\omega 
\end{bmatrix}
\begin{pmatrix}
\tilde{x}\\
\tilde{y}\\
\tilde{X}\\
\tilde{Y}
\end{pmatrix}
\nonumber \\
=
\begin{pmatrix}
\tilde{f}_{x}\\
\tilde{f}_{y}\\
0 \\
GH(\omega)\left(\cos\theta \tilde{f}_{x} + \sin\theta \tilde{f}_{y}\right),
\end{pmatrix}
\end{gather}
where the variables with the tilde signs represent the Fourier transforms of the variables without the tildes.

The solution of this system is readily found to be given by
\begin{gather}
\label{eq:X-solution}
\tilde{X}(\omega) = \mathrm{D}^{-1}(\omega)F(\omega), \\
F(\omega)=\left[ \mathrm{M}_x(\omega) \tilde{f}_{x} + \mathrm{M}_y(\omega) \tilde{f}_{y} \right] \nonumber,
\end{gather}
where the deterministic dynamics is defined by
\begin{gather}
\label{eq:D-def}
\mathrm{D}(\omega) = \omega_R^2 - \omega^2  \nonumber\\
-2g\omega_R	\frac{2g\delta + 2\kappa GH(\omega)\left[\delta\cos\theta + \left(\kappa + i\omega\right)\sin\theta\right]}{\delta^2 + \left(\kappa+i\omega\right)^2},
\end{gather}
while the noise contribution is mediated via
\begin{gather}
\label{eq:Mx-def}
\mathrm{M}_{x}(\omega) = \omega_R GH(\omega)\cos\theta \nonumber\\
+ \omega_R \frac{-2g(\kappa+i\omega) + 2\kappa GH(\omega) \left[\delta\sin\theta - \left(\kappa + i\omega\right)\cos\theta\right]}{\delta^2 + \left(\kappa+i\omega\right)^2} ,
\end{gather}
and
\begin{gather}
\label{eq:My-def}
\mathrm{M}_{y}(\omega) = \omega_R GH(\omega)\sin\theta \nonumber\\ 
- \omega_R \frac{2g\delta + 2\kappa GH(\omega) \left[\delta\cos\theta + \left(\kappa + i\omega\right)\sin\theta\right]}{\delta^2 + \left(\kappa+i\omega\right)^2} .
\end{gather}

The noise spectral correlation function $S(\omega)$ is defined via $\langle F(\omega) F(\omega')\rangle = S(\omega) \delta(\omega+\omega')$ and reads
\begin{equation}
\label{eq:S(omega)}
S(\omega) = \pi\kappa\left| \mathrm{M}_{x}(\omega) -i \mathrm{M}_{y}(\omega)\right|^2.
\end{equation}
This result is obtained assuming the fact that the feedback kernel $h(t)$ is a real function with the Fourier transform obeying $H(-\omega) = H^*(\omega)$.

In a more explicit form the noise spectral density reads:
\begin{gather}
\label{eq:S-explicit}
S(\omega) = \pi\kappa\omega_R^2\left| \frac{2g\left(\kappa + i(\omega-\delta)\right)}{\delta^2 + (\kappa+i\omega)^2}  \right. \nonumber\\
\left.+ GH(\omega)e^{-i\theta} \left(\frac{2\kappa \left(\kappa + i(\omega-\delta)\right)}{\delta^2 + (\kappa+i\omega)^2} - 1\right) \right|^2.
\end{gather}
The frequency dependence of $S(\omega)$ is due to both the effect of the feedback transfer function $H(\omega)$ and the dynamics of the cavity mode. As a result the noise acting on the atoms is non-Markovian even in the absence of feedback ($G = 0$). In the case of the large cavity decay rate $\kappa$ (large strength of the cavity measurement)  the only possible reason of non-Markovian dynamics is the feedback-induced noise processing. In the Fourier domain the adiabatic elimination of the cavity mode is formally done by neglecting $i\omega$ as compared with the decay rate $\kappa$. The adiabatic elimination results in the spectral correlation function presented in Ref. \cite{prl2020}:
\begin{equation}
\label{eq:S-adiabat}
S_a(\omega) = \frac{\pi \omega_R^2 \kappa}{\kappa^2 + \delta^2} \left|2g + G(\kappa - i \delta)e^{-i\theta}H(\omega)\right|^2.
\end{equation}

The critical value of the feedback gain in linear regime $G_\mathrm{crit}$ is determined from the condition $D(0) = 0$. If this condition is fulfilled then the low-frequency (long wavelength) fluctuations $\langle X^2\rangle$ will dominate and grow to infinity at the critical point. This is exactly what is expected at the critical point. Since the low-frequency range determines the critical value, the result for $G_\mathrm{crit}$ is exactly the same as that without the adiabatic elimination of the cavity field \cite{prl2020}:
\begin{equation}
G_\mathrm{crit} H(0) = \frac{\omega_R(\kappa^2 + \delta^2) - 4g^2\delta}{4g\kappa C_\theta}.
\end{equation}  
Note that this result exactly coincides with the critical value estimation (\ref{eq:G-crit-nl}) from the dynamical equations. 

The dynamics of the quadrature $X$ can also be formulated in time domain. The equation of motion reads
\begin{equation}
\label{eq:X-time}
\ddot{X}(t) + \omega_{R}^2 X(t) - 4 \omega_{R} g \int_{0}^{t}E^{\theta} \left(t -z\right) X(z)dz = F^{\theta}(t).
\end{equation}
Here the kernel $E^{\theta}$ is defined as
\begin{gather}
\label{eq:Es-theta}
E^{\theta}(t) = g e^{-\kappa	t} \sin \delta t  \nonumber\\
+G\kappa\int_{0}^{t} h(t-z) e^{-\kappa z} \sin \left(\delta z + \theta\right) dz,
\end{gather}
and the Langevin force operator appearing in the right-hand side of Eq.~(\ref{eq:X-time}) is expressed as
\begin{gather}
\label{eq:Langevin-force-time}
F^{\theta}(t) = -\omega_{R}\int_{0}^{t} dt' f_{a}(t')\left[g e^{-(\kappa + i\delta)(t-t')} \right. \nonumber\\
 - \frac{G}{2}e^{-i\theta}h(t-t')  \nonumber\\
\left.+ G\kappa e^{-i\theta}\int_{0}^{t-t'} dt'' h(t-t'-t'') e^{-(\kappa+ i\delta)t''} \right] + \mathrm{H. c.}
\end{gather}
In this form Eq.~(\ref{eq:X-time}) is similar to the equation describing Brownian motion in a harmonic potential \cite{Petruccione}. 

This integro-differential equation is similar to the one presented in Ref. \cite{prl2020} for the adiabatically eliminated light field, though with a very different kernel function. An important consequence of such a difference of kernels, consists in the fact that here (in the full model without the adiabatically eliminated light), the measurement of different quadratures (different $\theta$) leads to rather different equations. In contrast, in the simplified adiabatic model \cite{prl2020}, the measurement of different quadratures affects only some coefficients of the final simplified model.

The Heisenberg-Langevin approach allows for the calculation of the stationary fluctuations below the phase transition. The atomic $X$-quadrature fluctuation can be found from the correlation function
\begin{equation}
\label{eq:X2-SS}
\langle X^2 \rangle = \langle X(t+\tau) X(t) \rangle|_{\tau = 0} = \frac{1}{4 \pi^2} \int_{-\infty}^{\infty} \frac{S(\omega)}{|D(\omega)|^2} d\omega .
\end{equation} 
For the linear model this quantity becomes infinite as the feedback strength approaches the critical value $G_\mathrm{crit}$. 

We can now analyze the behavior of the quadrature fluctuation $\langle X^2 \rangle$ near the phase transition and evaluate the critical exponent $\alpha$. The dependence of $\langle X^2 \rangle$ on the feedback gain $G$ for $G<G_\mathrm{crit}$ will be numerically approximated by $\langle X^2 \rangle = A/|1-G/G_\mathrm{crit} |^\alpha + B$, where $A$ and $B$ are constants determined along with the critical exponent $\alpha$ during the least-mean-square fitting.

\section{Critical exponents}

\begin{figure*}
\centering
\includegraphics[width=0.99\linewidth]{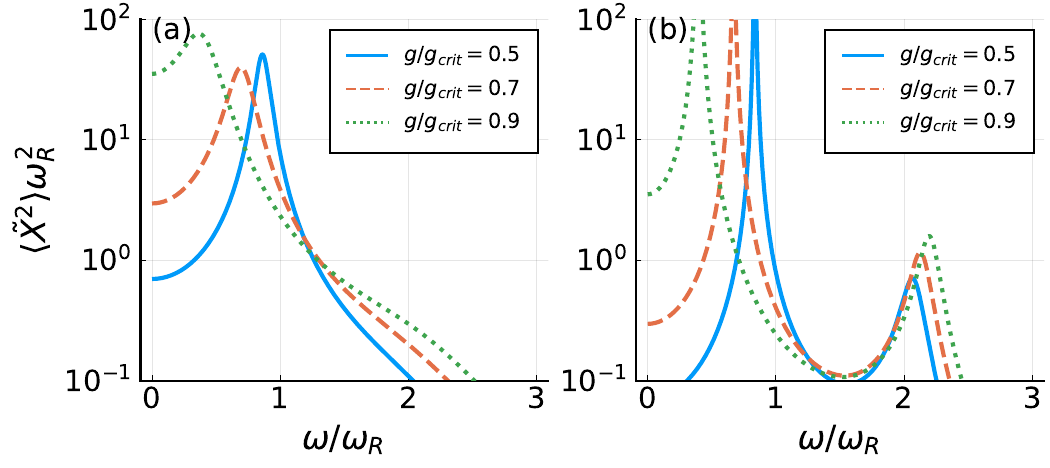}
\caption{\label{fig:no-fb-X2-spectrum}The quadrature fluctuation spectra for the system without feedback for different values of the atom--cavity coupling $g$. All spectra show the strong low frequency peaks that tend to the zero frequency, when one approaches the critical coupling $g_\mathrm{crit}$ (the so-called mode softening). (a) and (b) correspond to $\kappa = \omega_{R}$ and $\kappa = 0.1\omega_{R}$, respectively. In the case of lower damping (b), the quadrature fluctuations have another higher frequency peak that is only slightly shifted up during the increase of the coupling $g$. $\delta = 2\omega_R$.} 
\end{figure*}

In Ref. \cite{prl2020}, we have shown that a very useful choice of the feedback transfer kernel $h(t)$ is the power function of the form
\begin{eqnarray}\label{eq:h-def}
h(t)=h(0)\left(\frac{t_0}{t+t_0}\right)^{s+1} .
\end{eqnarray}
In view of the space--time analogy \cite{prl2020} this time dependence corresponds to the spatial Ising-type interaction with variable interaction length extensively used in different models of condensed matter physics. In addition, this system may correspond to models with the spatial long-range tunneling as well, where different critical exponents were found \cite{muller2021measurementinduced, zhang2021universal, block2021measurementinduced, minato2021fate}.

The instantaneous feedback ($s \rightarrow \infty$) with $h(t)\sim\delta(t)$ will lead to the short-range in time $S_x^2$ term as in the Lipkin-Meshkov-Glick (LMG) model \cite{Parkins2008,Muniz2020} originating from nuclear physics. A more exotic $h(t)$ such as a sequence of amplitude-shaped time delays $h(t)\sim \sum_n \delta(t-nT)/n^{s+1}$ will enable studies of discrete time crystals \cite{Sacha2017,UedaTC2018} and Floquet engineering \cite{Eckardt2017} with long-range interaction $\sum_n S_x(t)S_x(t-nT)/n^{s+1}$, where the crystal period may be $T=2\pi/\omega_R$. The global interaction is given by constant $h(t)$. The Dicke model can be restored even in the adiabatic limit by exponentially decaying and oscillating $h(t)$ mimicking a cavity response. 

We believe that our proposal will extend the studies in the field of time crystals \cite{Sacha2017,UedaTC2018,Demler2019,JakschNCom2019,JakschPRA2019,CosmePRA2019, KesslerNJP2020, kesler2020observation} and Floquet engineering \cite{Eckardt2017}. The time crystals is a recently proposed notion, where phenomena studied previously in space (e.g. spin chains, etc.) are now studied in the time dimension as well. Typically, the system is subject to the external periodic modulation of a parameter [e.g. periodic $g(t)$], which is considered as creating a ``lattice in time''. Our approach makes possible introducing the effective ``interaction in time.'' This makes the modulation in the system not prescribed, but dependent on the state of the system (via $S_x$). This resembles a true lattice in space with the interaction between particles. In other words, our model enables not only creating a ``lattice in time,'' but introducing the tunable ``interaction in time'' to such a lattice (without the necessity of having the standard particle-particle interaction in space \cite{FisherPhysRevLett.29.917,LuijtenPhysRevB.56.8945}).

Moreover, $h(t)$ can have minima, maxima, and even change its sign, which creates analogies with the molecular potentials (now in time, not in space) and rises intriguing questions about the creation of time molecules and time-molecule crystals.

 All such $h(t)$ can be realized separately or simultaneously to observe the competition between different interaction types. It will be interesting to extend this approach to a broader range of effective spin interactions \cite{Norcia259,DavisPRL2019,Muniz2020}. 
 
Further results of this paper do not rely on the effective Hamiltonians \cite{Caballero2015a}. This discussion motivates us to use in the simulations $h(t)$ given by Eq. (\ref{eq:h-def}), which is unusual in feedback control. In the numerical examples we will take $h(0)=s$ to make the critical value $G_\mathrm{crit}$ insensitive to the parameter $s$: this will allow us to focus on the influence of the feedback exponent $s$ on the quantum fluctuations, rather than on a trivial shift of the transition point.

Before addressing the results with feedback let us recall the behavior of the quadrature fluctuations when the system approaches the Dicke phase transition. The quadrature fluctuation spectra for this case are shown in Fig.~\ref{fig:no-fb-X2-spectrum}. Fig. \ref{fig:no-fb-X2-spectrum}(a) corresponds to $\kappa = \omega_{R}$ while Fig. \ref{fig:no-fb-X2-spectrum}(b) shows the results for $\kappa = 0.1\omega_{R}$. Different curves in both subplots represent different values of the coupling constant $g$. All spectra demonstrate strong peaks near the resonance frequency $\omega_{R}$ that move towards $\omega = 0$ as the coupling constant approaches its critical value, Eq.(\ref{eq:critical}). This is a known ``mode softening'' effect. 

In Fig.\ref{fig:no-fb-X2-spectrum}(b) the quadrature fluctuations have another weaker peak near the cavity resonance frequency $\delta = 2\omega_{R}$. It is the higher frequency mode of two coupled oscillators, atoms and the cavity field, that typically corresponds to the relative motion of the oscillators. This resonance is only slightly shifted to higher frequency as the coupling grows. Thus the oscillations at the frequency close to $\delta$ remain in the system even above the phase transition. The ``relative motion'' mode is also present in Fig.\ref{fig:no-fb-X2-spectrum}(a), but it is not resolved since the resonances for $\kappa = \omega_{R}$ are quite broad.

\begin{figure*}
\centering
\includegraphics[width=0.99\linewidth]{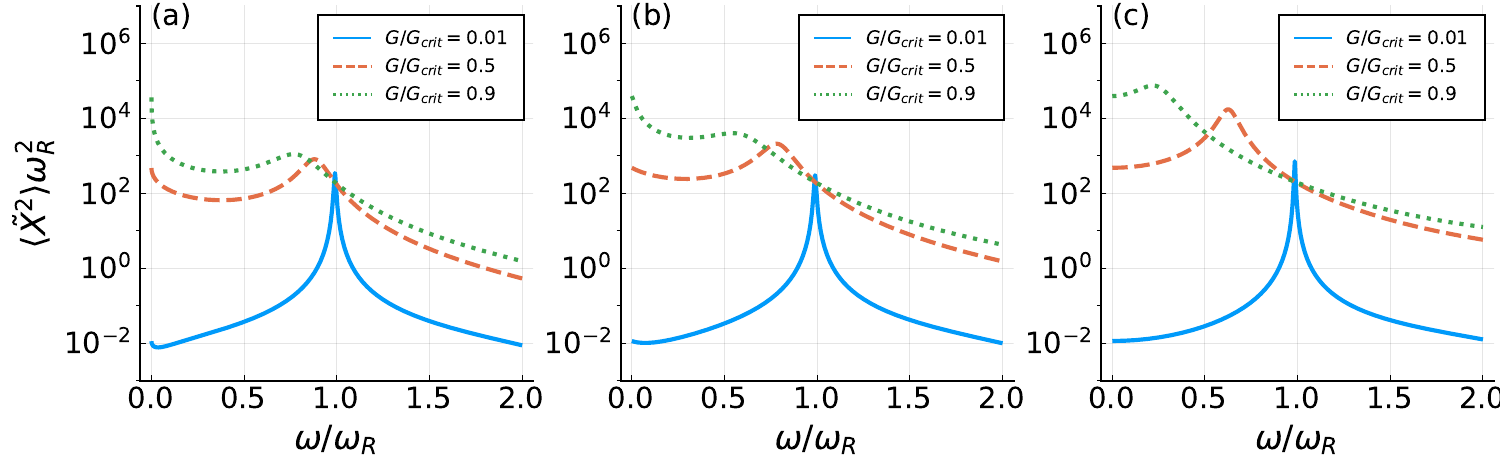}
\caption{\label{fig:fb-X2-spectrum}The quadrature fluctuations with feedback. (a) The slow feedback $s = 0.5$, (b) the results for $s = 1$, (c) the fast feedback $s = 5$. The systems far from the critical point are represented by solid curves, the systems close to the transition are represented by dotted curves, the intermediate cases are shown by dashed curves. While the fast feedback ($s = 5$) leads to a typical scenario of the mode softening (cf. Fig. \ref{fig:no-fb-X2-spectrum} with no feedback), the curves for slower feedbacks with $s = 0.5$ and $s = 1$ show a different behavior. $g = 0.1\omega_R$, $\kappa=\omega_R$, $\delta = 2\omega_R$, $\theta=\pi/2$, $h(0)=s$.}
\end{figure*}

The quadrature fluctuations with feedback are shown in Fig.~\ref{fig:fb-X2-spectrum}. Fig.~\ref{fig:fb-X2-spectrum}(a) corresponds to the slow feedback $s = 0.5$, Fig.~\ref{fig:fb-X2-spectrum}(b) shows the results for $s = 1$, and Fig.~\ref{fig:fb-X2-spectrum}(c) corresponds to the fast feedback $s = 5$. The systems far from the critical point are represented by solid curves, the systems close to the transition are represented by dotted curves, the intermediate case is shown by dashes.

The slow feedback, $s = 0.5$, demonstrates the trend that is quite different from the system without the feedback [compare Fig.\ref{fig:fb-X2-spectrum}(a) with Fig.~\ref{fig:no-fb-X2-spectrum}]. Increasing the gain $G$ the resonance near $\omega_{R}$ does not visibly move towards the zero frequency. Approaching the phase transition the fluctuations at zero frequency grow resulting in two pronounced peaks: around zero and around $\omega_{R}$.  The feedback frequency transfer function for the slow feedback has a narrow peak only at the zero frequency, thus the feedback does not strongly affect the shift of the resonance at   $\omega_{R}$. Therefore, the description of the system behavior in terms of the mode softening is not strictly valid for this case.

In Fig.~\ref{fig:fb-X2-spectrum}(b) the fluctuation spectra correspond to $s=1$. Here the shift of the resonance to lower frequency is more pronounced than for $s=0.5$, but the growth of the additional peak at zero frequency dominates.

The results for the fast, $s = 5$, feedback are shown in Fig.~\ref{fig:fb-X2-spectrum}(c). The transformation of the spectrum approaching the critical point is now quite similar to the transformation without the feedback (compare with Fig.~\ref{fig:no-fb-X2-spectrum}). The feedback transfer function is now broad enough to affect the system resonance frequencies. The action of feedback is similar to the increase of the coupling constant $g$. Thus one can expect that the critical behavior for the fast feedback (which is close to the instantaneous, Markovian, one) is similar to an open Dicke model without feedback. Note, that in the model with adiabatically eliminated light of Ref. \cite{prl2020} such a similarity is even stronger, because, as we have already mentioned, here the full model is more sensitive to the choice of the quadrature, which is detected, than the simplified model \cite{prl2020}.

\begin{figure*}
\centering
\includegraphics[width=0.99\linewidth]{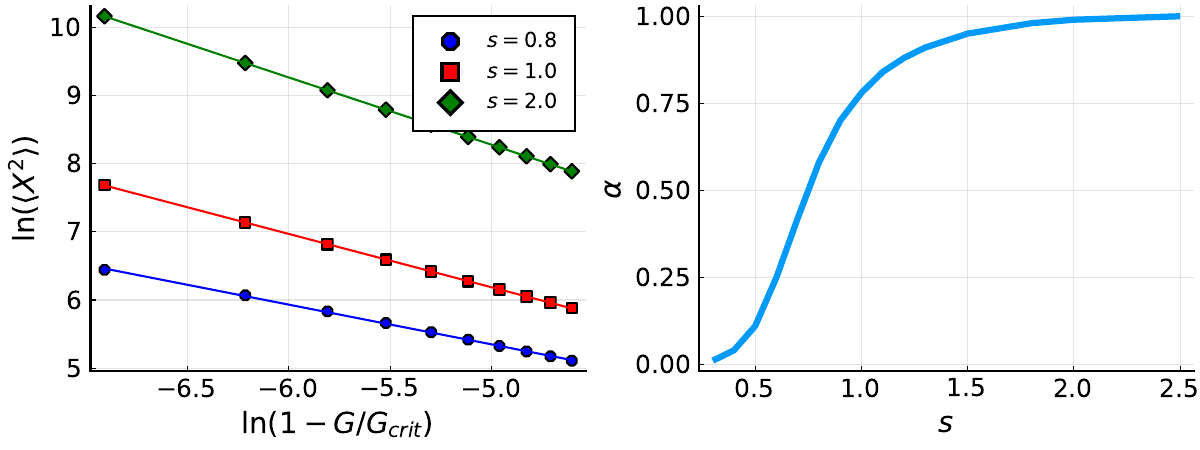}
\caption{\label{fig:fig3prl}Tuning the universality class of phase transitions. Integrated quadrature fluctuations $\langle X^2 \rangle$ (a) for different values of $s$ that are used to calculate the critical exponents and the dependence of critical exponent $\alpha$ on the feedback exponent $s$ (b). The lines in (a) show the fits for numerical simulations. The  growth of the quadrature fluctuations approaching $G_\mathrm{crit}$ (shown in logarithmic scale) is the manifestation of the phase transition in a linear model. The feedback time response determined by $s$ can continuously change the critical exponent of the transition in a wide range, which corresponds to tuning the universality class of the phase transition. $g = 0.5\omega_R$, $\kappa=\omega_R$, $\delta = 2\omega_R$,  $\theta=\pi/2$,  $h(0)=s$.}
\end{figure*}

Integrating the quadrature fluctuation spectra, Eq. (\ref{eq:X-solution}), similar to those presented in Fig.~\ref{fig:fb-X2-spectrum}, we obtain the average quadrature fluctuations below the phase transition. The examples of the fluctuation trends as the system approaches the critical point for different $s$ are shown in Fig.~\ref{fig:fig3prl}(a). The unlimited growth of the quadrature fluctuations approaching $G_\mathrm{crit}$ is the manifestation of the phase transition in a linear model. The growing rates for different $s$ looks different.

The dependencies presented in  Fig.~\ref{fig:fig3prl}(a) are approximated by the function $\langle X^2 \rangle = A/|1-G/G_\mathrm{crit} |^\alpha + B$ for $G<G_\mathrm{crit}$ fitting the parameters with the least-mean-square method. The results for the critical exponent $\alpha$ are calculated for different values of $s$ and are presented in Fig.~\ref{fig:fig3prl}(b). For small $s$ the critical exponent tends to zero. Increasing $s$ the critical exponent grows approaching the limiting value $\alpha = 1$. Thus the critical behavior of the system can be controlled by selecting the feedback time response. Choosing the value of $s$ one can therefore choose the universality class the system belongs to. This is very simple compared to the reservoir engineering approach.

According to the space--time analogy, varying the parameter $s$ of feedback response $h(t)$ (\ref{eq:h-def}) corresponds to varying the length of effective spin-spin interaction. For $h(t)$  (\ref{eq:h-def}), its spectrum is expressed via the exponential integral $H(\omega)=h(0)t_0e^{-i\omega t_0}E_{s+1}(i\omega t_0)$. At small frequencies its imaginary part behaves as $\omega^s$ for $s<1$, resembling the spectral function of sub-Ohmic baths. For large $s$, $\alpha$ approaches unity, as $h(t)$ becomes fast and feedback becomes nearly instantaneous  such as interactions in the open LMG and Dicke models, where $\alpha=1$ \cite{DomokosPRL2015, DomokosPRA2016,Oztop2012}.

\section{Quantum trajectories}

One can get an interesting insight in the system dynamics simulating single quantum trajectories. Weak measurements constitute a source of competition with unitary dynamics \cite{Mazzucchi2016PRA,Mazzucchi2016NJP, Mazzucchi2016SciRep,Kozlowski2016PRAnH}, which is well seen in quantum trajectories formalism \cite{Daley,Ruostekoski2014, Pedersen2014, Molmer2016PRA, VasilyevPRL2018, VasilyevPRA2018, Sherson2018PRA}, underlining the distinction between measurements and dissipation. Thus they can affect phase transitions, including the many-body ones \cite{Mazzucchi2016PRA, Kozlowski2016PRAnH, UedaCrit2016, AshidaNComm2017, AshidaAdvPhys2021,FisherPRB2019,PhysRevX.11.011030, muller2021measurementinduced, zhang2021universal, block2021measurementinduced, minato2021fate, doggen2021generalized, PhysRevLett.126.170602, PhysRevB.103.224210, PhysRevX.9.031009,PhysRevB.98.205136, PhysRevX.10.041020, PhysRevLett.125.030505, PhysRevB.101.104302, PRXQuantum.2.010352, PhysRevB.102.054302, Lavasani}.

The evolution of the quantum state of the continuously observed system is determined by the stochastic master equation 
\begin{equation}
\label{eq:SME}
d\rho_c = -i \left[H, \rho_c\right] dt + \mathcal{D}[a]\rho_c dt + \mathcal{H}[a]\rho_c dW,
\end{equation}
where the dissipation and noise contributions are defined as 
\begin{gather}
\label{eq:DH-def}
\mathcal{D}[a]\rho_c = 2\kappa \left[a\rho_c a^\dagger - \left(a^\dagger a \rho_c + \rho_c a^\dagger a \right)/2\right] , \nonumber\\
\mathcal{H}[a]\rho_c  = \sqrt{2\kappa} \left[ ae^{-i\theta} \rho_c + \rho_c a^\dagger e^{i\theta} 
\right. \nonumber\\
\left. - \mathrm{Tr}\left\{ae^{-i\theta}\rho_c + \rho_c a^\dagger e^{i\theta}\right\} \rho_c\right] .
\end{gather}
The increment of the Wiener process representing the quantum noise of the continuous weak measurement is denoted by $dW$. The Hamiltonian $H = H_s + H_{fb}$ is now the sum of the system Hamiltonian $H_s$, which is given by Eq.~(\ref{eq:H-BEC-Dicke}) for the nonlinear and by Eq.~(\ref{eq:H-BEC-Dicke-lin}) for the linearized spin models, and the feedback Hamiltonian is
\begin{equation}
\label{eq:H-fb}
H_{fb} = G \left(b^\dagger + b\right) I_c(t).
\end{equation}
Here $I_c$ is the feedback signal expressed similarly to Eq.~(\ref{eq:quadrature}), but it is not an operator:
\begin{equation}
\label{eq:I_c-def}
I_c(t) = \sqrt{2\kappa} \int_{-\infty}^{\infty} h(t-z) d\xi_\theta(z) .
\end{equation} 
The increment of the measured signal $d\xi_c$ being the outcome of a homodyne detector reads
\begin{equation}
\label{{eq:measured-S}}
d \xi_\theta(t) = \sqrt{2\kappa} \langle x_\theta \rangle_c dt + dW.
\end{equation}

The simulation of the stochastic evolution according to Eq. (\ref{eq:SME}) with the feedback kernels $h(t)$ as defined in Eq.~(\ref{eq:h-def}) will be presented for a single spin nonlinear model and $N$-spins linearized model.

\subsection{Quantum trajectories for linear model}

\begin{figure*}
\centering
\includegraphics[width=0.99\linewidth]{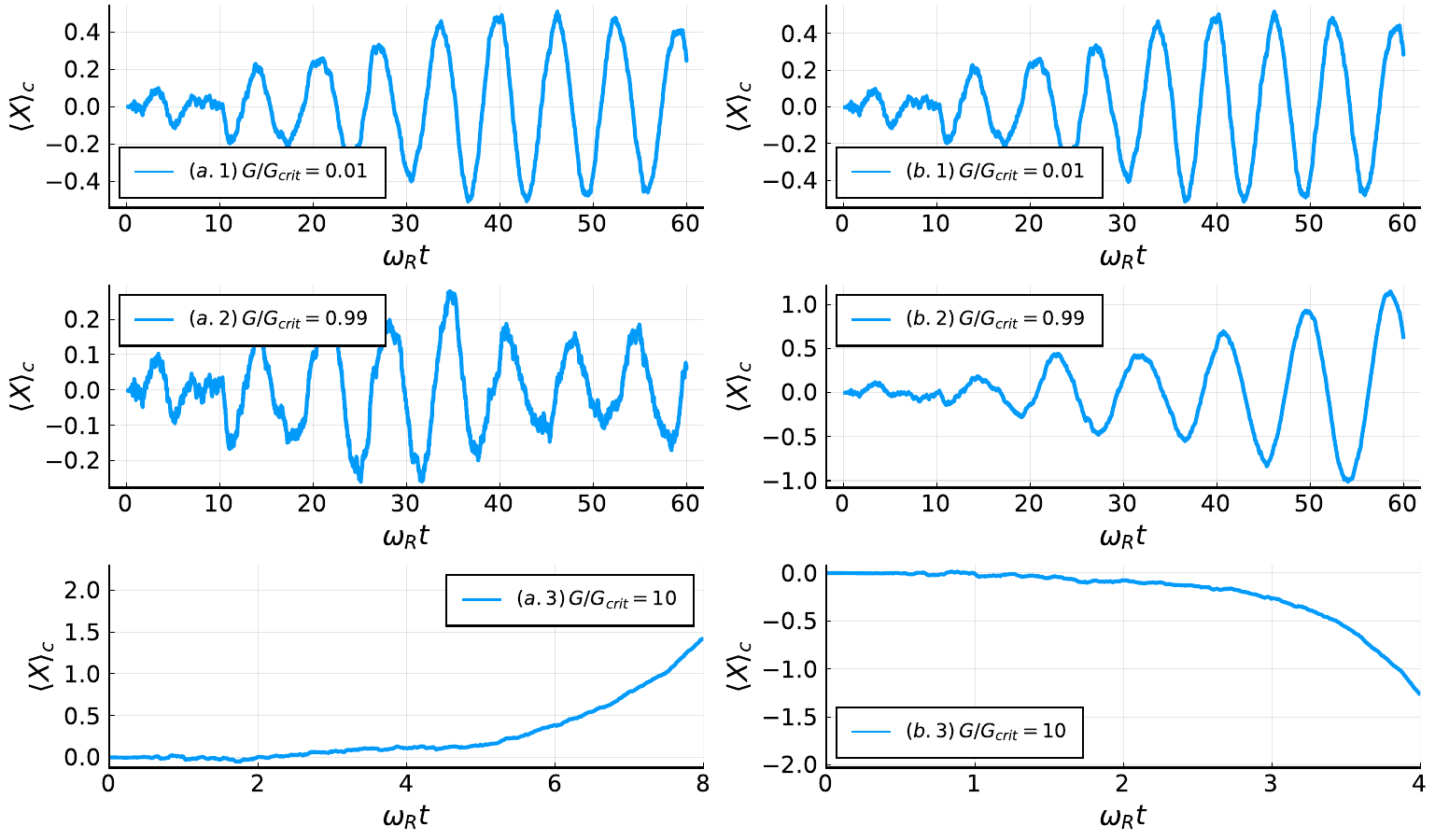}
\caption{\label{fig:lin-trj}Stochastic trajectories of the collective spin quadrature $\langle X \rangle_c$ for $s = 0.5$ (left column) and $s = 5$ (right column) in the linear approximation. Different rows correspond to different values of the feedback gain. For $G = 0.01 G_\mathrm{crit}$ the quadrature has an oscillating non-zero value on a single trajectory. Crossing the critical point $G_\mathrm{crit}$ the oscillatory dynamics changes to growth. The growth rate is grater for grater $s$. The different temporal behaviors for differend $s$ can be compared to the different fluctuation spectra for different $s$ in Fig. \ref{fig:fb-X2-spectrum}. One can see that the mode softening picture is not applicable to the slow feedback case: instead of decreasing frequency, one sees the increasing noise due to the appearance of the peak at zero frequency in Fig. \ref{fig:fb-X2-spectrum}(a). Other parameters are $g = 0.1\omega_{R}$, $\kappa = \omega_{R}$, $\delta = 2\omega_{R}$,  $\theta=\pi/2$,  $h(0)=s$.} 
\end{figure*}

Although we have the quasi-analytical result for the linearized spin model it is interesting to look at the behavior of the system at a single trajectory. The results of numerical simulations of Eq.~(\ref{eq:SME}) with the linearized Hamiltonian~(\ref{eq:H-BEC-Dicke-lin}) are shown in Fig.~\ref{fig:lin-trj}. Here the stochastic trajectories of collective spin quadrature $\langle X \rangle_c$ for $s = 0.5$ (left column) and $s = 5$ (right column) are shown for different values of $G$ (different rows).

If $G = 0.01G_\mathrm{crit}$ (the first row in Fig~\ref{fig:lin-trj}) the spin quadrature demonstrates quite regular oscillations. These oscillations are induced by the measurement backaction noise. They correspond to the resonant peaks in the fluctuation spectra in Figs.~\ref{fig:no-fb-X2-spectrum} and \ref{fig:fb-X2-spectrum}. 

Increasing $G$ and approaching its critical value we expect to observe the increase of the oscillation period reflecting the appearance of the soft mode. This is seen for the fast feedback with $s=5$ (right column in Fig. \ref{fig:lin-trj}), but this is not the case for the slow feedback with $s = 0.5$ (left column). For $G = 0.99G_\mathrm{crit}$ the quadrature oscillations for $s = 0.5$ become affected by the growing fluctuations at the zero frequency [see discussions of Fig.~\ref{fig:fb-X2-spectrum}(a) with spectra], while the spectral peak near the resonance frequency $\omega_R$ is still pronounced. This makes the oscillations noisy but still visible at the original frequency, which is different from the mode softening scenario.   

Above the critical point the oscillations completely disappear and the exponential growth of the spin quadrature takes place for both values of $s$, see the lowest row in Fig.~\ref{fig:lin-trj}. Note that growth rate is quite different for the slow and fast feedback. The time scale for the slow feedback is  larger than that for the fast one. Nevertheless, the phase transition takes place for both cases at the same value of $G_\mathrm{crit}$ (if we normalize the feedback kernel as $h(0)=s$).       

\subsection{Quantum trajectories for nonlinear single-spin model}

\begin{figure*}
\centering
\includegraphics[width=0.99\linewidth]{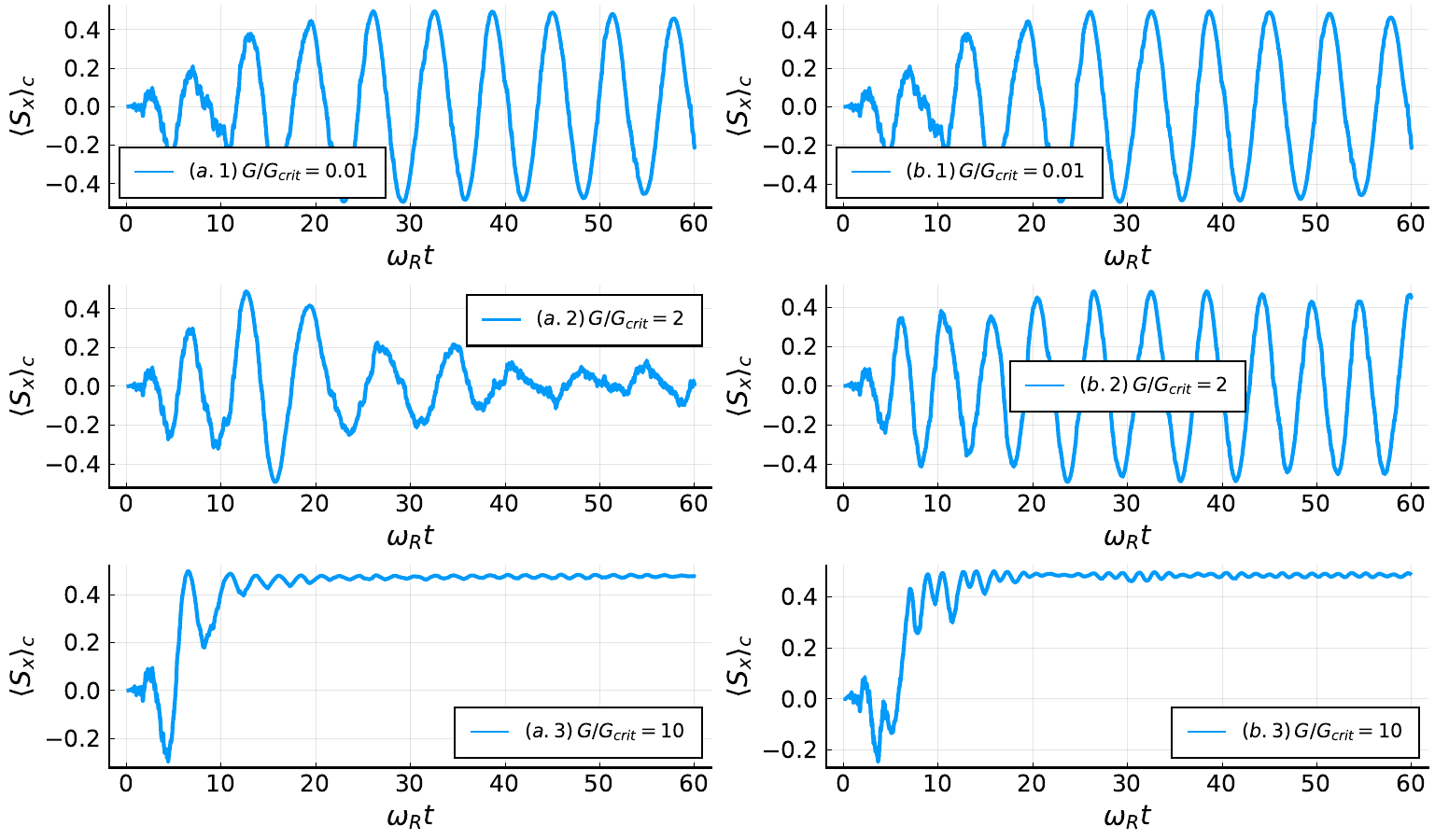}
\caption{\label{fig:nl_traj}Quantum trajectories for the conditional spin component $\langle S_x\rangle_c$ of a single nonlinear spin. The plots on the left-hand side, labeled (a), contain the trajectories for the slow feedback, $s = 0.5$. The right-hand side plots, labeled (b), show the fast feedback trajectories, $s = 5$. For the weak feedback ($G = 0.01G_\mathrm{crit}$) the spin starts to almost regularly oscillate due to the homodyne measurement backaction. There is no difference between $s=0.5$ (a.1) and $s=5$ (b.1).  Just above the threshold, $G = 2G_\mathrm{crit}$, the  feedback still results in oscillatory dynamics, which may be quite irregular (a.2). Well above the threshold $\langle S_x\rangle_c$ reaches a stationary value [(a.3) and (b.3)]. Other parameters are $g = 0.1\omega_{R}$, $\delta = 2\omega_{R}$, $\kappa = \omega_{R}$,  $\theta=\pi/2$,   $h(0)=s$.}
\end{figure*}

The examples of stochastic trajectories for a single nonlinear spin are shown in Fig.~\ref{fig:nl_traj}. It demonstrates the spin component $\langle S_x \rangle_c = \mathrm{Tr}\left\{\rho_c S_x\right\}$ averaged over the conditional state of the observed quantum system. The plots on the left-hand side, labeled (a), contain the trajectories for slow feedback, $s = 0.5$. The right-hand side plots, labeled (b), show the fast feedback trajectories, $s = 5$. For the weak feedback ($G = 0.01G_\mathrm{crit}$) the spin starts to almost regularly oscillate due to the homodyne measurement backaction. The oscillations are around zero value of $S_x$,  thus averaged over time such a behavior results in almost zero value of the spin component. This dynamics corresponds to the normal phase of the spin system and the cases of slow and fast feedbacks are very similar to each other. 

Just above the threshold, $G = 2G_\mathrm{crit}$, the feedback still results in oscillatory dynamics, which may be quite irregular, cf.  Fig.~\ref{fig:nl_traj}(a.2). This is similar to the behavior of the quadrature $X$ in the linearized model, cf. Fig.~\ref{fig:lin-trj}(a.2). Thus the slow feedback with the narrow band transfer function supports the low-frequency noise and suppresses oscillations [cf. Fig. \ref{fig:fb-X2-spectrum}(a) for the spectra of the linearized model]. 

\begin{figure}
\centering
\includegraphics[width=0.99\linewidth]{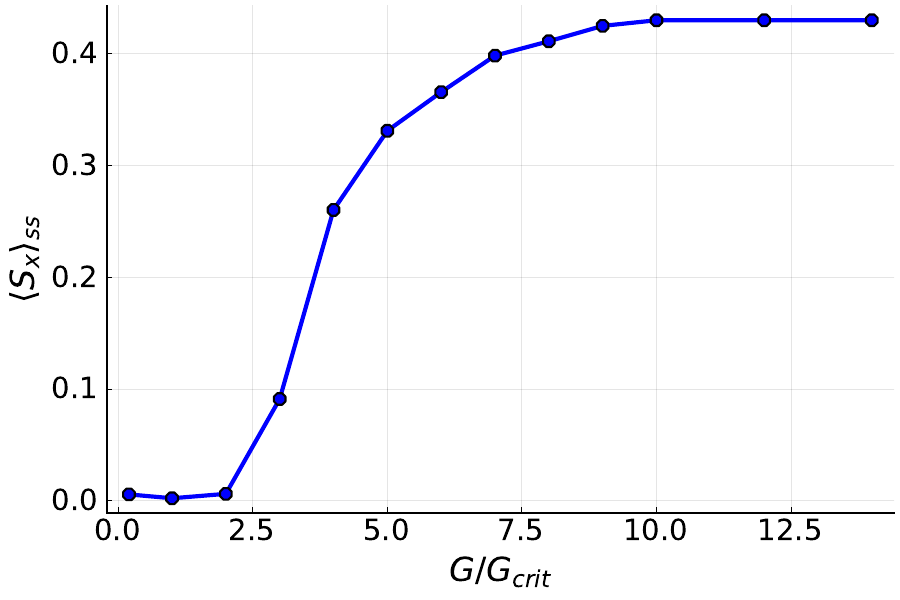}
\caption{\label{fig:nlgdep}Feedback-induced phase transition for a single nonlinear spin. The stationary values of $\langle S_x \rangle$ averaged over many trajectories depending on the feedback gain $G$.  $g = 0.1\omega_{R}$, $\kappa = \omega_{R}$, $\delta = 2\omega_{R}$, $s=1$,  $\theta=\pi/2$, $h(0)=s$.}
\end{figure}

Well above the threshold $\langle S_x \rangle_c$ rapidly saturates to its maximal value [Fig.~\ref{fig:nl_traj}(a.3) and (b.3)], but the sign of the spin component is not predefined and can take both positive and negative values. The dependence of the absolute value of such steady states averaged over many trajectories on the feedback parameter $G$ is shown in Fig. ~\ref{fig:nlgdep}. Below the critical point $G<G_\mathrm{crit}$ the spin component is nearly zero while it starts to grow for $G>G_\mathrm{crit}$. The stationary value of spin component saturates in the superradiant phase well above the threshold.

\section{Conclusions}

In summary, we presented a theoretical description of the feedback-induced quantum phase transition in a collective spin system and tuning its universality class. This shifts the paradigm of the feedback control from the control of quantum states (as known in quantum metrology) to the control of phase transitions in quantum systems. The feedback loop is assumed to be based on the quadrature measurement of the cavity light interacting with the spins. Starting with a nonlinear model we determined the critical value of the feedback gain and demonstrate that the dependence of the critical gain on the cavity decay rate is different for different measurement scenarios. The measurement of $x$-quadrature results in the critical gain linearly increasing with the decay rate, while the $y$-quadrature measurement results in the limited critical gain.

The analysis of the linearized spin model in the Heisenberg-Langevin approach allowed for the determination of the feedback-induced drift and the quantum noise acting on the system. The quantum noise is shown to be colored with the spectral density dependent on the feedback response shape resulting in the non-Markovian behavior of spins.

By approximating the trend of the spin-quadrature fluctuations near the critical gain by a power function, the critical exponents $\alpha$ have been found. The value of the critical exponent can be controlled in the range $\alpha \in (0,1)$ by changing the feedback loop timing parameter $s$. This opens new perspective for quantum simulations of systems belonging to different universality classes.

The dynamics of a single nonlinear spin and linearized $N$-spin models have been numerically simulated at  single quantum trajectories. For the feedback gain far below the critical value the main effect of the feedback is the measurement backaction. The quadrature measurement results in breaking the time symmetry and the appearance of coherent oscillations at a single quantum trajectory. If one calculates the expectation value of the spin variables, such oscillations are lost due to averaging over many trajectories, because of the random phase of the single-trajectory oscillations. Thus, the mean solution below the threshold is a trivial zero, while, in contrast, the coherent oscillations are visible in each single realization. Only for a fast (nearly instantaneous) feedback, approaching the critical point the period of the oscillations increases reflecting the softening of the spin mode. For the slow (long-memory) feedback such a simplified picture is not relevant.

While we considered both linear and nonlinear quantum systems, the feedback was assumed to be a linear one. A possible extension of the model can be the consideration of a feedback, nonlinearly dependent on the measured light variable.  This can be useful, e.g., for simulations of the spin-boson and other impurity models with the nonlinear coupling between the particle and bath \cite{ZhengPRB2018}. Such models can describe realistic systems such as the quadratic coupling of a qubit to its environment \cite{BertetPRL2005}, as well as experiments with semiconductor quantum dots \cite{PetersonPRL2010} and bismuth donors in silicon \cite{Wolfowicz}. 

Another possible extension of the model is considering several feedback loops simultaneously. Such a configuration can be useful for modeling spins coupled to several baths \cite{VojtaPRL2012}, which is important for simulating realistic qubits \cite{CastroPRL2003,NovaisPRB2005} and impurities in quantum magnets \cite{ZhuPRB2002,PhysRevB.66.024427,CastroPRL2003,NovaisPRB2005}.

\begin{acknowledgments}
The financial support is provided by RSF 17-19-01097-P (St. Petersburg State University, AMO physics approach), and DGAPA-UNAM IN109619 and CONACYT- 364 Mexico A1-S-30934 (Universidad Nacional Autónoma de México, CMT approach).
\end{acknowledgments}

\bibliographystyle{apsrev4-2}

%

\end{document}